\documentclass[prl,twocolumn,superscriptaddress,citeautoscript,showpacs,amsart]{revtex4}

\usepackage{float}
\usepackage{graphicx}
\usepackage{multirow}
\usepackage{color}
\usepackage{bm}
\usepackage{times}
\usepackage{amsmath,bm,amsfonts}
\usepackage{dcolumn}
\usepackage{graphicx}
\usepackage{latexsym}

\usepackage{ulem} 

\begin{document}

\title{Classification of accidental band crossings and emergent semimetals in two-dimensional noncentrosymmetric systems}

\author{Sungjoon \surname{Park}}
\affiliation{Department of Physics and Astronomy, Seoul National University, Seoul 08826, Korea}

\affiliation{Center for Correlated Electron Systems, Institute for Basic Science (IBS), Seoul 08826, Korea}

\affiliation{Center for Theoretical Physics (CTP), Seoul National University, Seoul 08826, Korea}

\author{Bohm-Jung \surname{Yang}}
\email{bjyang@snu.ac.kr}
\affiliation{Department of Physics and Astronomy, Seoul National University, Seoul 08826, Korea}

\affiliation{Center for Correlated Electron Systems, Institute for Basic Science (IBS), Seoul 08826, Korea}

\affiliation{Center for Theoretical Physics (CTP), Seoul National University, Seoul 08826, Korea}

\date{\today}

\begin{abstract}
We classify all possible gap-closing procedures which can be achieved in two-dimensional time-reversal invariant noncentrosymmetric systems. For exhaustive classification, we examine the space group symmetries of all 49 layer groups lacking inversion taking into account spin-orbit coupling. 
Although a direct transition between two insulators is generally predicted to occur when a band crossing happens at a general point in the Brillouin zone, we find that a variety of stable semimetal phases with point or line nodes can also arise due to the band crossing in the presence of additional crystalline symmetries. 
Through our theoretical study, we provide, for the first time, the complete list of nodal semimetals created by a band inversion in two-dimensional noncentrosymmetric systems with time-reversal invariance.
The transition from an insulator to a nodal semimetal can be grouped into three classes depending on the crystalline symmetry. Firstly, in systems with a two-fold rotation about the z-axis (normal to the system), a band inversion at a generic point generates a two-dimensional Weyl semimetal with point nodes.  Secondly, when the band crossing happens on the line invariant under a two-fold rotation (mirror) symmetry with the rotation (normal) axis lying in the two-dimensional plane, a Weyl semimetal with point nodes can also be obtained.
Finally, when the system has a mirror symmetry about the plane embracing the whole system, a semimetal with nodal lines can be created.
Applying our theoretical framework, we identify various two-dimensional materials as candidate systems in which stable nodal semimetal phases can be induced via doping, applying electric field, or strain-engineering, etc.
\end{abstract}

\pacs{}

\maketitle
 
{\it Introduction.$-$}
Recent discovery of three-dimensional (3D) Dirac \cite{theory_Young1, theory_Young2, LDA_Wang1, LDA_Wang2, Yang_classification1, Yang_classification2, Na3Bi_Shen, Yang_classification3, Cd3As2_Hasan, Cd3As2_Cava, Cd3As2_Yazdani, Cd3As2_Chen} and Weyl \cite{Weyl1, Weyl2, Weyl3, Weyl4, Weyl5, Weyl6, Weyl7, Weyl8, Weyl9} fermions in condensed matters has triggered intensive research in semimetals with point or line nodes, dubbed nodal semimetals (NSM).
Broadly, NSMs can be grouped into two classes.
In the first class, the degeneracy at the band crossing point/line is enforced by the nonsymmorphic space group symmetry of the system. 
In this class of NSMs, a certain minimal number of bands are required to stick together. Thus the presence of nodal points/lines at the Fermi level can be guaranteed by the electron filling~\cite{WPVZ}. 
On the other hand, in the second class of NSMs,
the gap-closing points/lines are created via a band inversion, that is, through a transition from an insulator to a semimetal via an accidental band crossing (ABC).
In this class of NSMs, the location of nodal points/lines in the momentum space varies depending on external parameters such as pressure, chemical doping, etc.
Here each nodal point/line carries a quantized topological charge, which guarantees the stability of NSMs~\cite{Yang_classification2,Yang_classification3, Murakami2, Murakami3}. In the case of semimetals with point nodes belonging to this class, a pair-creation/pair-annihilation of nodal points can even mediate topological quantum phase transitions between two insulators~\cite{Yang_classification1, Murakami2, Murakami3}.

In contrast to 3D, it is generally more difficult to have stable NSMs in two-dimensions (2D) due to the lower dimensionality. For instance, even the well-known Dirac fermions in graphene become unstable, and thus gapped, once {spin-orbit coupling (SOC) is included~\cite{Kane-Mele1,Kane-Mele2}. Recently, several interesting ideas have been proposed to stabilize 2D NSMs by using nonsymmorphic crystalline symmetries, thus leading to symmetry-enforced NSMs~\cite{Young-Kane,Wieder-Kane}.
However, there has been no systematic study yet on the other class of 2D NSMs created via a band inversion. Considering that the band gap of 2D systems is easier to control than that of 3D systems via gating or strain-engineering, it is essential to understand the outcome of a band inversion and the nature of resulting NSMs for future device application as well as for its fundamental physical aspect.

\begin{figure}[b]
\centering
\includegraphics[width=8.5 cm]{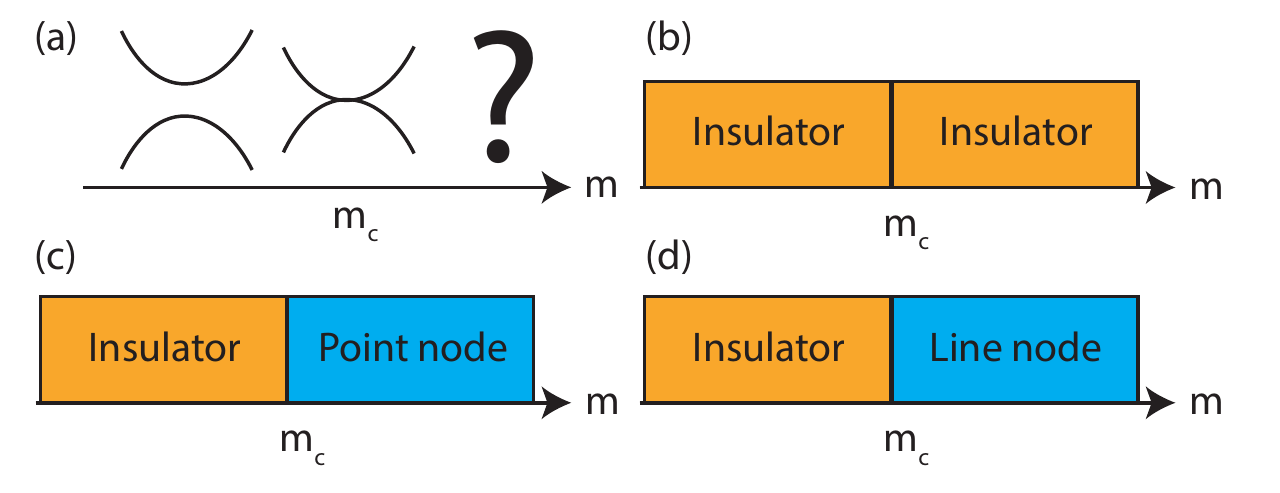}
\caption{
(a) Possible phase diagrams achieved by an ABC
in time-reversal invariant 2D noncentrosymmetric systems.
(b) An insulator-to-insulator transition when there is no additional symmetry at the gap-closing point. (c) Transition from an insulator to a semimetal with point nodes occurring when the system has either a two-fold rotation $C_{2x,2y,2z}$ or a mirror $M_{x,y}$. (d) Transition from an insulator to a semimetal with line nodes occurring when the system has a mirror $M_{z}$. 
}
\label{fig:phasediagram}
\end{figure}

In this letter, we classify all possible ABC events in time-reversal invariant 2D noncentrosymmetric systems for the first time. For exhaustive investigation of ABCs and the resultant semimetals, we use group theoretical approach by considering all possible layer groups (LGs) with broken inversion symmetry including SOC. We have found that there are three different types of ABC events as summarized in Fig.~\ref{fig:phasediagram}. 
In the first type, there is a direct transition between two insulators. In the second type, a band inversion creates a 2D Weyl semimetal with point nodes, i.e., Weyl points (WPs). Finally, in the third type, a nodal line semimetal is created by a band inversion.
At the critical point between an insulator and its neighboring phase, one can find characteristic fermionic excitations which lead to novel quantum critical behaviors. 
We propose various 2D materials in which our theory can be tested by engineering the electronic band structure .

{\it Classification of gap-closing events in layer groups.-}
Our strategy for classification of gap-closing events is as follows.
In the absence of inversion symmetry, energy bands are generally non-degenerate at a generic momentum $\bm{k}$. Thus, the relevant symmetry group at $\bm{k}$, the $\bm{k}$-group hereafter, would have a one-dimensional irreducible representation (1D irrep). In such cases,
the gap-closing at $\bm{k}_{0}$ between two nondegenerate bands can be described by a 2$\times$2 Hamiltonian 
\begin{equation}
H(\textbf{q},m)=f_{0}(\bm{q},m)+\sum_{i=1,2,3}{f_i(\bm{q},m)}\sigma_i,
\label{eqn:Hamiltonian}
\end{equation}
where $\sigma_i$ are the Pauli matrices describing the two bands and $f_{0,1,2,3}$ are real functions of the momentum $\bm{q}=\bm{k}-\bm{k}_{0}$ and an external parameter $m$ representing pressure, doping, etc. Here one can ignore $f_{0}$ as it does not contribute to the band gap. 
On the other hand, as discussed more fully in the Supplemental Materials, one may expect a band degeneracy associated with a higher dimensional irrep at some high symmetry lines or points like a time-reversal invariant momentum (TRIM)~\footnote{Because degeneracy due to time reversal symmetry along high symmetry points or lines is not tabulated for layer groups in~\cite{Litvin}, we have analysed the issue in detail in the Supplemental Materials}.
However, since the bands degenerate at $\bm{k}_{0}$ generally disperse linearly away from $\bm{k}_{0}$, the band minimum or maximum is located away from $\bm{k}_{0}$, which means that an ABC always happens away from $\bm{k}_{0}$. Thus, we can limit ourselves to the case where the irrep of the conduction and the valence bands, $R_c$ and $R_v$, respectively, are one-dimensional with the effective Hamiltonian in Eq.~(\ref{eqn:Hamiltonian})~\cite{Murakami1}.
Since the symmetry of a 2D crystal embedded in a 3D space is described by a layer group (LG), one can exhaustively classify all possible gap-closing events in 2D by analysing the 49 inversion asymmetric LGs in the presence of SOC.

\begin{figure}[t]
\centering
\includegraphics[width=8 cm]{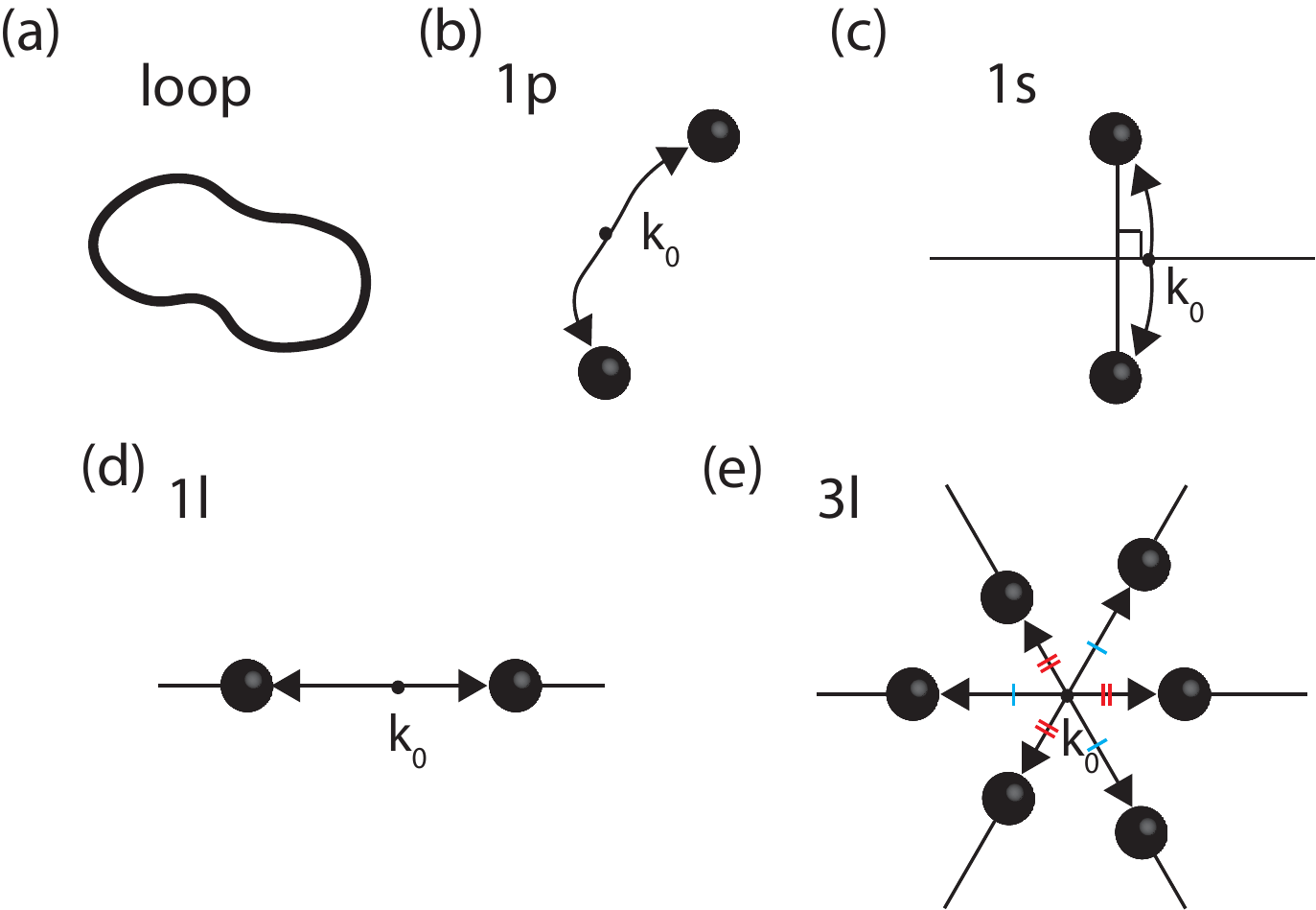}
\caption{Schematic figures describing the structure of gap-closing points
created by a band inversion in 2D momentum space. 
(a) \textbf{loop}: A line node moves in the plane.
(b) \textbf{1p}: A WP pair moves in the plane.  
(c) \textbf{1s}: A WP pair moves symmetrically with respect to a symmetry line. 
(d) \textbf{1l}: A WP pair moves along a symmetry line. 
(e) \textbf{3l}: Three pairs of WPs move along symmetry lines.
}
\label{patterns}
\end{figure}

Suppose that the band gap of a system which can be tuned by varying $m$ stays finite for $m<m_c$ but closes at $m=m_c$. We are interested in the nature of this system when $m>m_c$.
To describe a gap-closing at a generic momentum $\textbf{k}_0$, three equations $f_{1,2,3}=0$ must be satisfied. Since we have three parameters  $(k_{x},k_{y},m)$, we expect a unique solution near the critical point. Such a solution describes the critical point between two insulators, as illustrated in Fig.~\ref{fig:phasediagram}(b). However, when the $\bm{k}$-group at the gap-closing point $\bm{k}_{0}$ has certain crystalline symmetries that impose constraints on $f_{1,2,3}$, the gap-closing condition can be modified leading to NSM when $m>m_{c}$.
Below, we list all symmetries in a $\bm{k}$-group that give non-trivial solutions to the problem at hand. We work out the nonsymmorphic symmetry explicitly only in the case \textbf{[b]} below since a similar idea can be applied to the other cases.

\textbf{[a] No symmetry}: There is no constraint on $f_{1,2,3}$, thus one can find a unique gap-closing solution $(\bm{k}_{0},m_{c})$. In this case, an ABC occurs only through fine tuning. We label this process by \textbf{f} representing for fine tuning. 

\textbf{[b] Two-fold rotation $C_{2x}$ (similarly for $C_{2y}$ or mirror $M_{x}$, $M_{y}$)}: (\textbf{i})  If $R_c=R_v$, we may take $C_{2x}=c\sigma_0$ where $\sigma_0$ is the $2\times 2$ identity matrix and  $\left| c \right|=1$. Here $c$ may be a function of $\textbf{k}$ if we consider a nonsymmorphic counterpart of this symmetry. On a symmetry line, the Hamiltonian is as in Eq.~\eqref{eqn:Hamiltonian} since $C_{2x}$ does not give any further constraint. Thus, the gap closing condition gives 3 equations whereas there are only 2 variables, that is, $m$ and the momentum on the symmetry line. Thus, in general, the band-gap cannot be closed on the symmetry line. (\textbf{ii})  If $R_c=-R_v$, we can write $C_{2x}=c\sigma_3$ so $H=f_3 \sigma_3$ on the symmetry line. In this case, the gap-closing problem has two variables and one equation so the solution is one-dimensional in the parameter space. We label this as \textbf{1l} where \textbf{1} denotes the number of WP pairs created and \textbf{l} indicates the WPs are on the symmetry line. (See Fig.~\ref{patterns} (d).) 


\textbf{[c] $C_{2z}\Theta$}: $C_{2z}\Theta$ is a local symmetry in the 2D momentum space. Since $C_{2z}\Theta$ is anti-unitary, its general form is $C_{2z}\Theta=UK$ where $K$ denotes complex conjugation and $U$ indicates a unitary matrix. After a suitable unitary transformation, one can always have $C_{2z}\Theta=K$ as shown in Supplemental Materials. Since $C_{2z}\Theta$ requires the Hamiltonian $H(\bm{k})$ to be real, $f_2=0$. Then the gap closing condition gives 2 equations whereas there are 3 parameters. This means that the solution is one-dimensional, and this describes a creation of a WP pair and their evolution in the momentum space. We label this by \textbf{1p}, where \textbf{p} stands for the plane where WPs are located and \textbf{1} indicates the number of WP pairs. Let us note that we count the number of WP pairs locally. In fact, $C_{2z}$ implies that there is another WP pair created at $\bm{-k_{0}}$. (See Fig.~\ref{patterns} (b).)  
Let us also note that in systems with $C_{2z}\Theta$, the Weyl semimetal is stable irrespective of the eigenvalues of the bands since each WP carries a quantized $\pi$ Berry phase~\cite{Fang-Fu, Ahn}. 

\textbf{[d] $M_z$}: $M_{z}$ is also a local symmetry in the 2D momentum space. ($\textbf{i}$) If $R_c=R_v$, only fine tuning gives 2D WPs since there are three equations and three variables. ($\textbf{ii}$) If $R_c=-R_v$, one can choose $M_z=\sigma_3$ which gives $H=f_3\sigma_3$. The gap-closing condition gives one equation while we have three parameters, so ABC occurs in a 2D manifold in the parameter space, which translates to the creation of a line node and its evolution. Since the gap-closing points, in general, form a loop in the momentum space, we label it by $\textbf{loop}$. (See Fig.~\ref{patterns} (a).) 


\textbf{[e] $C_{2x}$ and $C_{2y}\Theta$ (similarly for $C_{2y}$ and $C_{2x}\Theta$, or$M_{x(y)}$ and $M_{y(x)}\Theta$)}: 
Since $C_{2x}C_{2y}\Theta\propto C_{2z}\Theta$ ($M_{x}M_{y}\Theta\propto C_{2z}\Theta$),
a WP is stable even when it is away from high-symmetry axes. 
($\textbf{i}$)
Considering $C_{2x}$ eigenvalues, if $R_c = R_v$, 
$C_{2x}=i\sigma_0$ and the Hamiltonian is not constrained by $C_{2x}$ on its invariant axis.
However, due to $C_{2z}\Theta$, the Hamiltonian should be real.
Then on the $C_{2x}$ invariant axis, the gap-closing condition gives 2 equations with two parameters including the momentum along the invariant axis and $m$, which leads to the case \textbf{f} on the invariant axis. However, more detailed analysis shows that the gap-closing on the $C_{2x}$ invariant axis creates a pair of WP that move symmetrically away from the invariant axis. We label this case as \textbf{1s} where \textbf{s} means symmetrical. (See Fig.~\ref{patterns} (c).) 
($\textbf{ii}$) If $R_c = -R_v$, one can choose $C_{2x}=i\sigma_3$. 
Then the Hamiltonian on the invariant axis depends only on $f_{3}$, and the gap-closing condition gives 1 equation with 2 parameters, which describes the creation of a WP pair following the pattern \textbf{1l}. 

\textbf{[f] $C_{2x}$ and $M_{y}$}: Since $C_{2x}M_{y}\propto M_{z}$, a nodal line can appear after a band inversion. Let us note that $C_{2x}$ and $M_{y}$ share the same invariant line. 
($\textbf{i}$) If $\{C_{2x},M_{y}\}=0$ on the invariant line (recall that these can be nonsymmorphic), two bands with different $C_{2x}$ (or $M_{y}$) eigenvalues are doubly degenerates. In this case, a band inversion does not happen on the invariant line. 
($\textbf{ii}$) If $[C_{2x},M_{y}]=0$ on the invariant line, each band on the invariant line carries $C_{2x}$ and $M_{y}$ eigenvalues simultaneously. When a band inversion happens between two bands with different $C_{2x}$ ($M_{y}$) eigenvalues while sharing the same $M_{y}$ ($C_{2x}$) eigenvalues, a nodal line is created after the band inversion corresponding to \textbf{loop}.
If both $C_{2x}$ and $M_{y}$ eigenvalues are different between two bands, the band inversion creates a WP pair on the invariant line corresponding to \textbf{1l}. 

\textbf{[g] $C_{3}$ plus $C_{2x}$ or $M_{y}$}: 
This happens at the $K$ or $KA$ point of the hexagonal Brillouin zone.
Since two bands with $C_{3}$ eigenvalues $e^{i\pi/3}$ and $e^{-i\pi/3}$, respectively,
are degenerate at $K$ or $KA$, a band inversion can happen only between two bands with $C_{3}$ eigenvalue -1. When these two bands carry different $C_{2x}$ or $M_{y}$ eigenvalues, a band inversion can happen and create 3 pairs of WPs, which are located 
on the lines invariant under $C_{2x}$ or $M_{y}$. We label it as \textbf{3l} as shown in Fig~\ref{patterns} (e).

{\it Classification table.$-$}
We summarize all possible gap-closing patterns in Table~\ref{results} for 49 LGs lacking inversion symmetry. 
In the first column of Table~\ref{results}, we list the group numbers for the inversion asymmetric LGs following the convention in Ref.~\cite{ITC}. In the second column, we list the corresponding space group number which is formed by stacking the layered system. The precise relation is that for each LG L, there is a space group G such that if T(1) is a one-dimensional translation subgroup, $L\approx G/T(1)$ \cite{Hitzer, Litvin}. In the third column, we list the possible gap closing patterns. Here we use the notation ii to mean $R_c=R_v$, and ij to mean $R_c\neq R_v$. We also use the notation ii:ij:\textbf{1s},\textbf{1l} to mean ii leads to \textbf{1s} and ij leads to \textbf{1l}. $\langle 4 \rangle$:\textbf{loop},\textbf{1l} is used for case \textbf{[f]} above, where there are 4 possible 1D irreps. In this case, different $M_z$ eigenvalues lead to \textbf{loop} while different $C_{2x}$ or $M_y$ eigenvalues lead to \textbf{1l}. In the case of \textbf{1p}, we do not specify $R_c$ and $R_v$ since WPs are stable independent of eigenvalue spectra. Here the labels on the Brillouin zone follow the conventions used in Ref.~\cite{Litvin}. For the reader's convenience, we have illustrated the Brillouin zones in the Supplemental Materials~\footnote{A caveat is in order: the group numbers used in \cite{ITC} differs from those used in \cite{Litvin}. The relation between these conventions can be found in \cite{Notation}}.

\begin{table}[t]
\centering
\caption{Classification table of all possible gap-closing patterns for 49 inversion asymmetric layer groups (LGs). The first column indicates the LG numbers used in Ref.~\cite{ITC}. The second column denotes the corresponding space group. When there are multiple groups sharing the same gap-closing pattern, the LG and the corresponding space group are listed in the same order. The third column describes gap-closing patterns.}
\label{results}
\begin{tabular}{|l|l|l|}
\hline
Layer group & Space group &Gap-closing pattern \\
\hline
\textbf{1}, \textbf{65}&\textbf{1}, \textbf{143}& \textbf{f} \\
\textbf{3}, \textbf{49}, \textbf{50}, \textbf{73}&\textbf{3}, \textbf{75}, \textbf{81}, \textbf{168}& \textbf{1p} \\
\textbf{4}, \textbf{5}, \textbf{27}, \textbf{28}, \textbf{29},&\textbf{6}, \textbf{7}, \textbf{25}, \textbf{26}, \textbf{26},& ij:\textbf{loop} \\
\textbf{30}, \textbf{35}, \textbf{36}, \textbf{74}, \textbf{78}&\textbf{27},\textbf{35}, \textbf{39}, \textbf{174}, \textbf{187}& ij:\textbf{loop} \\
\textbf{79}&\textbf{189}& ij:\textbf{loop} \\
\textbf{31}, \textbf{32}, \textbf{33}, \textbf{34}&\textbf{28}, \textbf{31}, \textbf{29}, \textbf{30}&ij:\textbf{loop}; $\langle 4 \rangle$:\textbf{loop},\textbf{1l} DA, D\\
\textbf{8}, \textbf{9}&\textbf{3}, \textbf{4}& ij:\textbf{1l} $\Delta$, TA, D, DA \\
\textbf{10}&\textbf{5}& ij:\textbf{1l} DA, $\Delta$, FA, F \\
\textbf{11}, \textbf{12},\textbf{13}&\textbf{6}, \textbf{7}, \textbf{8}& ij:\textbf{1l} $SN,\Sigma,CA,C$ \\
\textbf{19},  \textbf{23}&\textbf{16}, \textbf{25}& \textbf{1p}; ii,ij:\textbf{1s},\textbf{1l} $\Delta$, D, $\Sigma$, C \\
\textbf{20}& \textbf{17}& \textbf{1p}; ii,ij:\textbf{1s},\textbf{1l} $\Delta$, D, $\Sigma$\\
\textbf{24}&\textbf{28}&\textbf{1p}; ii,ij:\textbf{1s},\textbf{1l} $\Delta$, $\Sigma$, C \\
\textbf{21}, \textbf{25}, \textbf{54}, \textbf{56}, \textbf{58}& \textbf{18}, \textbf{32}, \textbf{90}, \textbf{100}, \textbf{113}& \textbf{1p}; ii,ij:\textbf{1s},\textbf{1l} $\Delta$, $\Sigma$\\
\textbf{60}&\textbf{117}&\textbf{1p}; ii,ij:\textbf{1s},\textbf{1l} $\Delta$, $\Sigma$\\
\textbf{22}, \textbf{26}&\textbf{21}, \textbf{35}& \textbf{1p}; ii,ij:\textbf{1s},\textbf{1l} $\Sigma$, $\Delta$, F, C \\
\textbf{53}, \textbf{55} \textbf{57}, \textbf{59}&\textbf{89}, \textbf{99}, \textbf{111}, \textbf{115}& \textbf{1p}; ii,ij:\textbf{1s},\textbf{1l} $\Delta$, $\Sigma$, Y \\
\textbf{67}, \textbf{69}&\textbf{149}, \textbf{156}& ij:\textbf{1l} $\Sigma$, SN \\
\textbf{68}, \textbf{70}&\textbf{150}, \textbf{157}& ij:\textbf{1l} $\Lambda$, T, TA, LE; ij:\textbf{3l} K, KA\\
\textbf{76}, \textbf{77}&\textbf{177}, \textbf{183}& \textbf{1p}; ii,ij:\textbf{1s},\textbf{1l} $\Lambda$, $\Sigma$, T; ij:\textbf{3l} K \\
\hline
\end{tabular}
\end{table}

{\it Effective Hamiltonian at the quantum critical point.$-$} 
To describe the effective Hamiltonian at the quantum critical point with $m=m_{c}$, we redefine the coordinates so that the gap closes at $\textbf{k}=\textbf{k}_0=0$ and $m=m_{c}=0$. As described below, the effective Hamiltonian at the critical point falls into three categories. 

Firstly, when there is an insulator-to-insulator transition,
the bands disperse linearly in two directions at the critical point as shown in Fig.~\ref{graphs}(b).
The relevant effective Hamiltonian is
$H=a_1k_1 \sigma_1+a_2k_2 \sigma_2$ where we use $k_1$, $k_2$ since they are not along $k_x$ and $k_y$ in general. Secondly, at the critical point where a pair of WPs is created, the bands disperse linearly in one direction but quadratically in the other direction. In particular, if the WPs are protected by $C_{2x,2y}$ or $M_{x,y}$, the relevant Hamiltonian is 
$H=a_1k_y\sigma_1+a_{3}k_x^2 \sigma_3$ (Fig.~\ref{graphs} (c)) whereas
in the case with $C_{2z}\Theta$, it is
$H=a_{1}k_1^2\sigma_1+(a_{2}k_1^2+a_3k_2)\sigma_3$.
Note that the presence of $k_1^2$ in the coefficient of $\sigma_3$ breaks $k_2\rightarrow -k_2$ symmetry.
Finally, there are two cases in which the bands disperse quadratically in two directions.
One is at the critical point between an insulator and a nodal line semimetal
with the Hamiltonian 
$H=(a_1 k_1^2+ a_2 k_2^2)\sigma_3$
where we require $a_1 a_2>0$.
The other case is at the critical point where three pairs of WPs are created (\textbf{3l}).
The relevant Hamiltonian is $H=u_{1}k^3\sin3\theta\sigma_1+u_{3r}k^2\sigma_{3}$
where $u_{1},~u_{3r}$ are constants and $k_{x}+k_{y}=ke^{i\theta}$.
(See the Supplemental Materials for the detailed form of the effective Hamiltonian covering $m<m_{c}$ and $m>m_{c}$ cases as well.)

\begin{figure}[t]
\centering
\includegraphics[width=8.5cm]{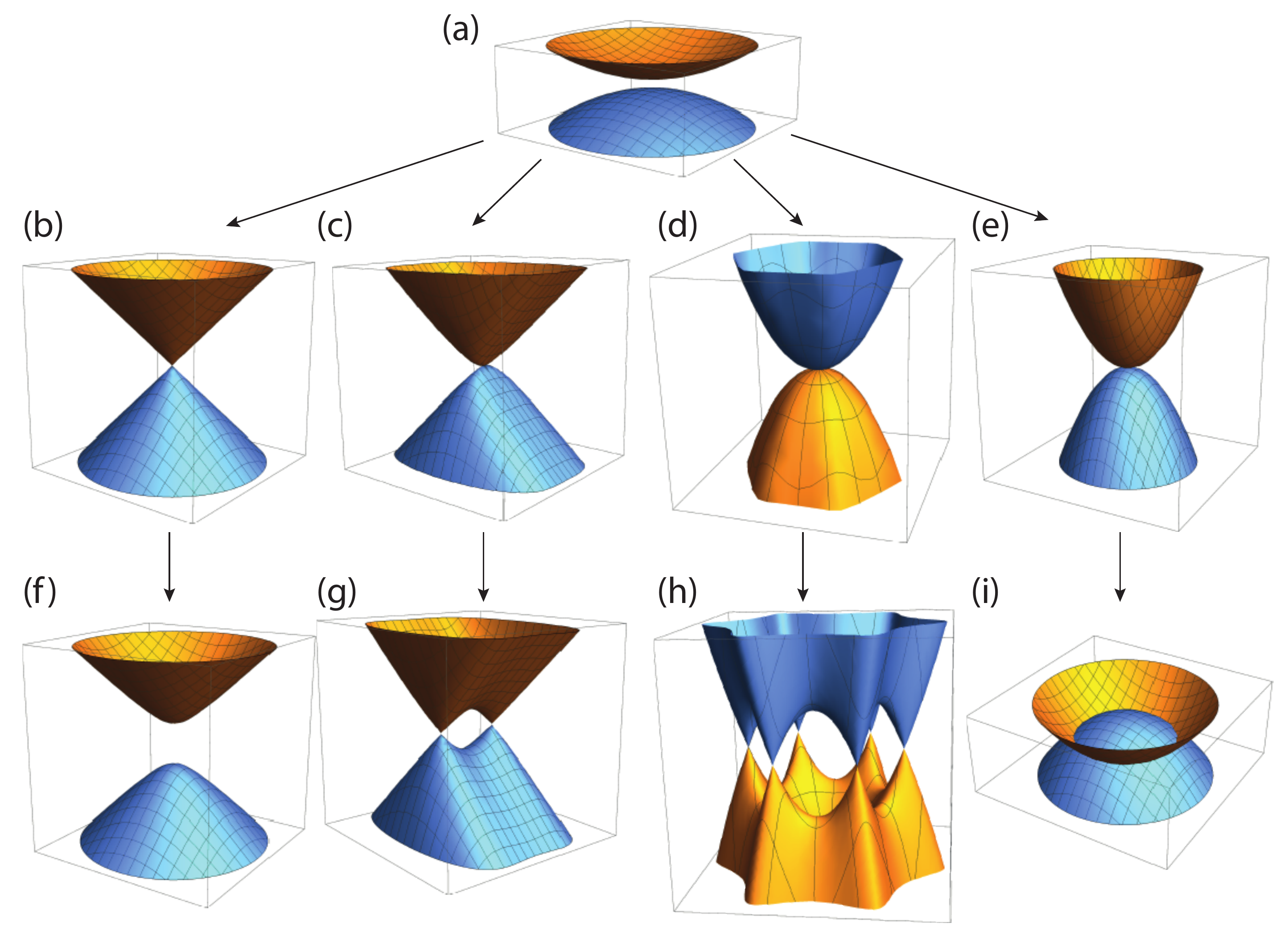}
\caption{Evolution of the band structure across an ABC. (a) Bands before the gap-closing with $m<m_{c}$. (b-e) Bands at the critical point with $m=m_{c}$. (f-i) Bands after the gap-closing with $m>m_{c}$. (b,f) For an insulator-to-insulator transition. (c,g) For a transition to a Weyl semimetal protected by $C_{2x,2y}$ or $M_{x,y}$ or $C_{2z}\Theta$. 
(d, h) For a transition to a Weyl semimetal protected by $C_{3}$ together with
$C_{2x,2y}$ or $M_{x,y}$. 
(e, i) For a transition to a nodal line semimetal protected by $M_z$.}
\label{graphs}
\end{figure}

{\it Application to 2D materials.$-$} 
Our theory can be applied to various 2D materials whose band gap is widely tunable by gating, doping, or strain-engineering. 
Let us first focus on the variants of the 2D planar honeycomb lattice since many 2D materials fall into this category. Because we have organized our results according to LG, it suffices to identify the LG of the lattice structure. The planar honeycomb lattice has the structure of the LG \textbf{80}. By distorting the lattice, it is possible to obtain a puckered structure belonging to the LG \textbf{42}, and buckled structure with the LG \textbf{69}~\cite{Kamal-Ezawa, Ahn, Sb, Bi, blue-phosphorous, SiGe}. Although the planar and the puckered structures contain inversion symmetry, 2D materials are usually fabricated on a substrate, and this breaks inversion symmetry (one could instead apply electric field normal to the plane of the material). 
Then, the symmetry of the planar and the puckered structure is lowered to LG \textbf{77} and LG \textbf{24}, respectively. Another variant of the honeycomb lattice structure is the dumbbell structure, whose symmetry group, like the planar structure, is also LG \textbf{80}~\cite{Sb, Bi}.
Of course, there are also 2D materials whose structure is not based on the honeycomb lattice. For instance, Bi$_4$Br$_4$ has the structure belonging to LG \textbf{18} which lowers to LG \textbf{13} upon breaking inversion symmetry ~\cite{Bi4Br4}. HgTe in HgTe/CdTe quantum well belongs to LG \textbf{57}\cite{Ahn}. We summarize the candidate systems and their LGs in Table~\ref{materials}. Once the LG for the given material is determined, all possible gap-closing patterns can be read off from TABLE~\ref{results}.

\begin{table}[t]
\centering
\caption{List of candidate 2D materials: The first column lists the lattice structure. The second column lists the layer group (LG) number for the structure considering inversion symmetry breaking effect. The third column lists specific materials that fall under the category.}
\label{materials}
\begin{tabular}{|c|c|c|}
\hline
Structure & ~LG~ & Material examples\\
\hline
Planar honeycomb & \textbf{77} & graphene \\
Puckered honeycomb& \textbf{24} &  arsenene \cite{Kamal-Ezawa}, antimony \cite{Sb},  \\
& & bismuth \cite{Bi}, black phosphorus \cite{Ahn}\\
Buckled honeycomb& \textbf{69} & arsenene \cite{Kamal-Ezawa}, blue phosphorous \cite{blue-phosphorous}, \\ 
& & silicon, germanium \cite{SiGe}, antimony \cite{Sb}, \\ & & bismuth \cite{Bi} \\
Dumbbell & \textbf{77} & stanene \cite{stanene}, Sn$_6$Ge$_4$, Sn$_6$Ge$_4$H$_4$ \cite{SnGe} \\
Bi$_4$Br$_4$ & \textbf{13} & Bi$_4$Br$_4$ \cite{Bi4Br4}\\
HgTe & \textbf{57} & HgTe/CdTe heterostructure \cite{Ahn}\\
\hline
\end{tabular}
\end{table}

{\it Discussion.$-$}
One important application of our classification table is to use it for searching
unconventional mechanisms for topological quantum phase transitions~\cite{Ahn}.
For instance, recent theoretical and experimental studies on few-layer black phosphorus
have shown that it is possible to achieve a transition from an insulator to a Weyl semimetal by doping potassium ions. Due to its puckered structure, few-layer black phosphorus under vertical electric field belongs to LG 24.
Since the gap-closing happens on the $k_{x}$ axis invariant under $M_{y}$, $C_{2z}\Theta$,
and the $M_{y}$ eigenvalues of the conduction and valence bands are identical (see Supplemental Materials), the gap-closing pattern should be \textbf{1s}, which is confirmed by theoretical studies.
Interestingly, such an emergent 2D Weyl semimetal phase can mediate a transition between a normal insulator and a quantum spin Hall insulator~\cite{Ahn}.
Considering that an insulator-to-insulator transition is generally predicted in 2D time-reversal invariant systems, the emergence of stable nodal semimetals through an ABC can provide a potential source to achieve unconventional topological phase transitions.

Moreover, the unusual fermion dispersion at the critical point between an insulator and a nodal semimetal can generate unconventional quantum critical phenomena. In particular, the density of states $D(E)$ at the energy $E$ shows $D(E)\propto E$ for the linear dispersion in two directions, $D(E)\propto \sqrt{E}$ for the linear-quadratic dispersion, and $D(E)\propto$ const. for the quadratic dispersion in two directions. Since the system is more susceptible to interaction or disorder as the low energy density of states increases, one can expect unconventional quantum critical behavior, in particular, when the dispersion is quadratic in two directions. In fact, previous theoretical studies on 2D semimetals with quadratic band crossing have shown the the short-range Coulomb interaction is marginally relevant. Thus, it can induce various insulating phases with broken symmetries.
Since such a quadratic dispersion is expected at the critical point in our problem, it is natural to expect novel quantum critical behavior associated with an ABC, which we leave for future study.

{\it Acknowledgement.$-$} 
S. Park was supported by IBS-R009-D1.
B.-J. Y was supported by IBS-R009-D1, Research Resettlement Fund for the new faculty of Seoul National University, and Basic Science Research Program through the National Research Foundation of Korea (NRF) funded by the Ministry of Education (Grant No. 0426-20150011).



\appendix
\setcounter{equation}{0}
\setcounter{figure}{0}
\setcounter{table}{0}
\setcounter{section}{1}
\setcounter{subsection}{1}
\renewcommand{\thesection}{S\Roman{section}}
\renewcommand{\thesubsection}{S\arabic{subsection}}
\renewcommand{\theequation}{S\arabic{equation}}
\renewcommand{\thefigure}{S\arabic{figure}}

\section{\thesection. General kp Hamiltonian}
\addtocounter{section}{1}
As explained in the main text, we may write
\begin{equation}
H=\sum_{i=1}^{3}{a_i \sigma_i},
\label{generic} 
\end{equation}
which we will use throughout this section (we have ignored a term proportional to the 2 by 2 identity matrix because it does not contribute to the band gap). Hermiticity of the Hamiltonian requires that and $\textit{\textbf{a}}=(a_1,a_2,a_3)$ be real functions of $k_1,~k_2,~m$. Let us redefine the coordinates so that the gap closes at $m=0$, $\textbf{k}=(k_1,k_2)=0$. Also define $\textbf{q}=(q_1,q_2,q_3)=(k_1,k_2,m)$. With this notation, the gap closes at $\textbf{q}=0$. We will also sometimes use $k_x$, $k_y$ instead of $k_1$, $k_2$ when it is more convenient to fix the direction of the coordinates. We will also frequently make use the following property of Pauli matrices. 

Suppose that $\sigma_i'=\sum_{j=1}^{3}{O_{ij}\sigma_j}$ is a set of matrices obtained from the Pauli matrices by an orthogonal transformation $\textbf{O}$. Then, \textit{if $H'=\sum_{i=1}^3{b_i'\sigma_i'}$, the eigenvalues are $E_{\pm}'=\pm\sqrt{b_1'^2+b_2'^2+b_3'^2}$}. To show this, note first that the eigenvalues of a Hamiltonian of the form \eqref{generic} are $E_{\pm}=\pm\sqrt{a_1^2+a_2^2+a_3^2}$. To see this, use the fact that Pauli matrices transform like a vector under $SU(2)$. Thus, there is always an $SU(2)$ transformation that takes the Hamiltonian to $H=\pm\sqrt{a_1^2+a_2^2+a_3^2}\sigma_3$, from which the statement follows. Now, using the fact that $\textbf{O}$ is an orthogonal matrix, $H'=\sum_{i=1}^{3}{b_i'\sigma_i'}=\sum_{i,j,k=1}^{3}{b_k' O_{ik} O_{ij} \sigma_j'}=\sum_{i=1}^{3}{b_i \sigma_i}$, where $b_i=\sum_{k=1}^{3}{O_{ik}b_k'}$ and $\sigma_i'=\sum_{j=1}^{3}{O_{ij}\sigma_j}$. Then, it follows that $E_{\pm}'=\pm\sqrt{b_1^2+b_2^2+b_3^2}=\pm\sqrt{b_1'^2+b_2'^2+b_3'^2}$. Thus, $\sigma_i'$ obtained from orthogonal transformation of the Pauli matrices are just as good for expanding the Hamiltonian.

\subsection{\thesubsection. \textbf{No Symmetry}}
\addtocounter{subsection}{1}

In this subsection, we explore in more detail how the gap closes for the case labelled by \textbf{f} in the main text. Expanding $\textit{\textbf{a}}$ to first order in $\textbf{q}$ around the gap closing point, we have $\textit{\textbf{a}}=\textbf{M}\textbf{q}$. Here, the matrix $\textbf{M}$ has components $M_{ij}=\frac{\partial a_i}{\partial q_j}$, $i,j=1,2,3$. We first examine what happens when $\textbf{M}$ is not invertible. If the matrix has rank 2, the solution is one dimensional in the parameter space while if the matrix has rank 1, it is two dimensional\footnote{Let $\textbf{M}$ be an m by n matrix. The rank of $\textbf{M}$ is the number of independent rows, which is equivalent to the number of independent columns. The nullity of $\textbf{M}$ is the dimension of the solution space of the linear equation $\textbf{M}\textbf{v}=0$. The rank-nullity theorem states that the rank of $\textbf{M}$ and the nullity of $\textbf{M}$ adds up to n.}. Thus, for these cases, a gap-closing solution exists for arbitrary value of $m$. Since we are assuming that the gap is open when $m<0$, these cases can be excluded from our consideration. Note that the case $\textbf{M}=0$ is unlikely. To see this, carry out the singular value decomposition of $\textbf{M}=\textbf{A}^T \textbf{D} \textbf{B}$, where $\textbf{A}$ and $\textbf{B}$ are orthogonal matrices while $\textbf{D}$ is diagonal. If $\textbf{M}$ is not invertible, one or more of the entries of $\textbf{D}$ is zero, which should not happen without special reason. Our constraint that there is no gap closing for $m<0$ while there is at least one gap closing point at $m=0$ does not give such a constraint, so we expect $\textbf{M}$ to be invertible, and in particular, $\textbf{M}\neq 0$.

Thus, \textbf{M} is in general invertible, and there is only one solution to the gap-closing condition in the neighborhood of $\textbf{q}=0$. This gives a Hamiltonian with linearly dispersing bands which are degenerate at $\textbf{k}=0$ when $m=0$ but with quadratically dispersing with a gap when $m \neq 0$. To see this, first write the Hamiltonian as
\begin{equation}
H=\sum_{i,j=1}^{3}{M_{ij}q_j\sigma_i}.
\end{equation}
Carry out the QR decomposition on the matrix $\textbf{M}=\textbf{Q}\textbf{R}$. Here, $\textbf{Q}$ is a orthogonal matrix and $\textbf{R}$ is an upper triangular matrix. Redefine $\sigma_i'=\sigma_j Q_{ji}$ and $q_i'=R_{ij}q_j$ so that $\sum_{ij}{M_{ij}q_j\sigma_i}=\sum_{i}{q_i' \sigma_i'}$. Notice that we may carry out the decomposition such that the diagonal components of $\textbf{R}$ are positive. This follows because $\det(\textbf{R})=R_{11}R_{22}R_{33}\neq 0$, and whenever any one of $R_{ii}$ ($i=1,2,3$) is negative, the sign may be absorbed into the matrix $\textbf{Q}$. For example, if $R_{11}$ is negative, define $\textbf{D}=diag(-1,1,1)$. Then, $\textbf{Q}\textbf{R}=\textbf{Q}\textbf{D}\textbf{D}\textbf{R}$. The QR decomposition can be carried out with $\textbf{Q}'=\textbf{Q}\textbf{D}$ and $\textbf{R}'=\textbf{D}\textbf{Q}$ instead, in which case $R_{11}'$ is positive. The $\sigma'_i$ are orthogonal transformation of the Pauli matrices and $\textbf{q}'=(k_1', k_2', m')$, where $k_1'$ and $k_2'$ are linear transformation of $k_1$ and $k_2$ while $m'=c m$ for a positive constant $c$. The Hamiltonian is then
\begin{equation}
H=k_1'\sigma_1'+k_2'\sigma_2'+m'\sigma_3'.
\end{equation}
Now, it is easier to see that the dispersion is linear when $m=cm'=0$ while the dispersion is quadratic when $m'\neq0$. Note, however, that this transformation comes with a price that $k_1'$ and $k_2'$ no longer forms an orthogonal coordinate system. The gap closing process is illustrated in FIG.~\ref{SMgraphs} (a), (b), (g).

When $m=0$, we can write $\textit{\textbf{a}}=\textbf{L}\textbf{k}$ where $\textbf{L}$ is the 3 by 2 matrix with components $L_{ij}=\frac{\partial a_i}{\partial k_j}$ ($i=1,2,3$ and $j=1,2$). Use the singular value decomposition on $\textbf{L}$ to write 
$\textit{\textbf{a}}=\textbf{U}^T \mathbf{\Sigma}\textbf{V} \textbf{k}$ where $\textbf{U}$ and $\textbf{V}$ are orthogonal matrices and $\mathbf{\Sigma}$ is a 3 by 2 rectangular diagonal matrix with the only nonzero entries $\Sigma_{11}=v_1$, $\Sigma_{22}=v_2$. Defining $\textit{\textbf{a}}'=\textbf{U}\textit{\textbf{a}}$ and $ \textbf{k}'=\textbf{V} \textbf{k}$, the Hamiltonian can be written as 
\begin{equation}
H=v_1k'_1 \sigma_1'+v_2k'_2 \sigma_2',
\end{equation}
where $\sigma_i'$ is orthogonal transformation of the Pauli matrices.

\subsection{\thesubsection. \label{S2} \textbf{$\mathbf{C_{2x}}$ or $\mathbf{M_{y}}$ Symmetry}} 
\addtocounter{subsection}{1}

In this subsection, we carry out a similar analysis for the case labelled \textbf{1l} in the main text. As explained in the main text, the requirement for stable band crossing is that  $R_c\neq R_v$, where $R_c$ and $R_v$ are the 1D irreducible representations of the symmetry for the conduction and the valence band along the high symmetry line. This restricts the Hamiltonian to $H=a_3 \sigma_3$  on the symmetry lines $k_y=0,~\pi$. Furthermore, off the symmetry axis, $a_{1,2}=k_{y}b_{1,2}(k_x,k_y^2)$ and $a_3=a_3(k_x,k_y^2)$ due to the constraint that $H(k_x,-k_y)=C_{2x}H(k_x,k_y)C_{2x}^{-1}$. We can approximate $a_3=a_{3x}k_x+a_{3yy}k_y^2+a_{3xx}k_x^2+a_{3m}m$. We must now implement the condition that there should be no solution for $m<0$ but that a solution exists for $m=0$ (along the symmetry line). Setting $k_y=0$ in $a_3$, $a_3=a_{3x}k_x+a_{3xx}k_x^2+a_{3m}m$. The number of solutions is determined by the discriminant, $D=a_{3x}^2-4a_{3xx}a_{3m}m$. Thus, we must have $a_{3x}=0$ while $a_{3xx}a_{3m}<0$. Now, make the following expansions: $a_{1,2}=k_{y}a_{1,2y}$, $a_3=a_{3xx}k_x^2+a_{3m}m$ (we do not include $k_x k_y$ because there is a term linear in $k_y$ that will overwhelm $k_xk_y$ when $k_y\neq 0$, while when $k_y=0$, it is zero. $mk_y$ was ignored for similar reasons). Thus, the effective Hamiltonian is
\begin{equation}
H=a_{1y}k_y\sigma_1+a_{2y}k_y\sigma_2+(a_{3xx}k_x^2+a_{3m}m)\sigma_3.
\end{equation}
This describes closing of the gap and the subsequent evolution of Weyl points as shown in FIG.~\ref{SMgraphs} (a), (c), (h).  When $m=0$, we have $H=k_ya_{1y}\sigma_1+k_ya_{2y}\sigma_2+a_{3xx}k_x^2\sigma_3$. Carrying out a rotation in the $\sigma_1$, $\sigma_2$ space, we find
\begin{equation}
H=a'_1k_y\sigma'_1+a_{3xx}k_x^2 \sigma_3.
\end{equation}
Here, the set $(\sigma'_1,\sigma_2',\sigma_3)$ is an orthogonal transformation of the Pauli matrices and $a'_1=\sqrt{a_{1y}^2+a_{2y}^2}$. Note that the dispersion is linear in $k_y$ direction but quadratic in $k_x^2$ direction (See FIG.~\ref{SMgraphs} (c)). Note also that Weyl points move in the quadratically dispersing direction, which is equivalent to the direction of the high symmetry line.

\subsection{\thesubsection. \textbf{Space Time Inversion}}
\addtocounter{subsection}{1}

In this subsection, we carry out a similar analysis for the case \textbf{1p} discussed in the main text. As explained in section SII S1, we can choose the basis so that the space time inversion symmetry (STI) $I_{ST}=C_{2z}\theta$ is represented as $K$, where $K$ is a complex conjugation operator. In such a basis, $a_2$ will vanish. Expanding about the gap closing point, $a_1=M_{1x}k_x+M_{1y}k_y+N_{1}m$, $a_3=M_{3x}k_x+M_{3y}k_y+N_{3}m$. The gap closing condition is that $\textit{\textbf{a}}=\textbf{M}\textbf{k}+m\textbf{N}=0$ (here and in the next subsection, $\textit{\textbf{a}}=(a_1,a_3))$) $M_{ij}=\frac{\partial a_i}{\partial k_j}$; $N_i=\frac{\partial a_i}{\partial m}$ where $i=1,3$ and $j=x,y$). If \textbf{M} is invertible, there would be a solution for arbitrary m, in contradiction to the assumption that there is no solution for $m<0$. Thus, $\det \textbf{M}=0$ and there exist $\textbf{n}_1$ such that $\textbf{M}\textbf{n}_1=0$. Choose $\textbf{n}_3$ orthogonal to $\textbf{n}_1$ and expand \textbf{k} using this basis: $\textbf{k}=k_1 \textbf{n}_1+k_2 \textbf{n}_3$. Defining $\textbf{u}_3=\textbf{M}\textbf{n}_3$, we have $\textit{\textbf{a}}=k_2\textbf{u}_3+m\textbf{N}$. There is no $k_1$ term so we must expand to higher orders. The lowest allowed $k_1$ term is $k_1^2$. We include only this term since other higher order terms will be overwhelmed away from $k_2=0$ by terms linear in $k_2$. Then the lowest order approximation is $\textit{\textbf{a}}=\textbf{u}_{11} k_1^2+\textbf{u}_3k_2+\textbf{N}m$.

Now, choose $\textbf{u}_1$ so that $\hat{\textbf{u}}_1$ and $\hat{\textbf{u}}_3=\textbf{u}_3/|\textbf{u}_3|$ form an orthonormal basis. Then, we may expand $\textit{\textbf{a}}$ in terms of this basis:
\begin{align}
\textit{\textbf{a}}=& \textbf{u}_{11} k_1^2+\textbf{u}_3k_2+m\textbf{N} \nonumber \\
=&((u_{11})_1\hat{\textbf{u}}_1+(u_{11})_3\hat{\textbf{u}}_3)k_1^2+u_3k_2\hat{\textbf{u}}_3+m(N_1'\hat{\textbf{u}}_1+N_3'\hat{\textbf{u}}_3)
\end{align}
The gap closing condition $\textit{\textbf{a}}=0$ can be written
\begin{equation}
0
=
\begin{pmatrix}
(u_{11})_1&0\\ 
(u_{11})_3&u_3
\end{pmatrix}
\begin{pmatrix}
k_1^2\\ 
k_2
\end{pmatrix}
+
m\begin{pmatrix}
N_{1}'\\ 
N_{3}'
\end{pmatrix}
=\textbf{U}\tilde{\textbf{k}}+m\textbf{N}.
\end{equation}

This can be solved for $\tilde{\textbf{k}}=(k_1^2, k_2)$ by inverting the matrix $\textbf{U}=(\textbf{u}_{11}, \textbf{u}_3)$: $\tilde{\textbf{k}}=-m\textbf{U}^{-1}\textbf{N}=m\textbf{Q}$. We require $Q_1=-N_1'/(u_{11})_1>0$ to get solution for $m\geq 0$. To get an expression for the Hamiltonian, notice that choosing $\hat{\textbf{u}}_1$ and $\hat{\textbf{u}}_3$ as the basis for expanding $\textit{\textbf{a}}$ constitutes a change of basis by an orthogonal matrix $\textbf{P}^T=(\hat{\textbf{u}}_1, \hat{\textbf{u}}_3)$. If we carry out a similar change of basis for the Pauli matrices to get $\sigma_i'$, the Hamiltonian can be written as $H=\sum_{i=1,3}{\sigma_ia_i}=\sum_{i,j,k=1,3}{\sigma_jP_{ij}P_{ik}a_k}=\sum_{i=1,3}{\sigma_i' a_i'}$. Here, $a_i'$ are components of $a_i$ in the new basis because $\textbf{P}\textit{\textbf{a}}=\textit{\textbf{a}}'$. Explicitly, 
\begin{equation}
H=((u_{11})_1k_1^2+N_1m)\sigma'_1+((u_{11})_3k_1^2+u_3k_2+N_2m)\sigma'_3.
\label{sti1}\end{equation}
This describes gap closing and evolution of Weyl points as shown in FIG.~\ref{SMgraphs} (a), (d), (i). For $m=$0, the Hamiltonian can be written as
\begin{equation}
H=(u_{11})_1k_1^2\sigma'_1+((u_{11})_3k_1^2+u_3k_2)\sigma'_3.
\label{sti2}\end{equation}
As in the previous case, the energy is linear in $k_2$ direction but quadratic in $k_1$ direction. However, this case is slightly different in that there is no $k_2 \rightarrow -k_2$ symmetry. The Weyl points move in the quadratically dispersing direction in this case as well. This can be seen from the gap closing conditions, which are $(u_{11})_1 k_1^2+N_1 m=0$ and $(u_{11})_3k_1^2+u_3k_2+N_2m$. The former shows that for $m$ slightly greater than 0, $k_1 \sim \sqrt{m}$ and the latter shows that $k_2 \sim m$. Thus, for small $m$, the Weyl points move predominantly in the quadratically dispersing direction.

\subsection{\thesubsection. \textbf{$\mathbf{C_{2x}}$, $\mathbf{C_{2z}\theta}$}}
\addtocounter{subsection}{1}

We expand on the discussion in the main text in a similar manner with the symmetries $C_{2x}$ and $C_{2z}\theta$, which give rise to either the pattern \textbf{1l} or the pattern \textbf{1s}. Since the case for \textbf{1l} was discussed in subsection SI S2, we discuss only the case \textbf{1s}. For this case, we may take $C_{2x}=i\sigma_0$ and $C_{2z}\theta=K$ (we assumed that the conduction and the valence bands both have eigenvalues $+i$ since the case for eigenvalues $-i$ is similar). These symmetries restrict the Hamiltonian to $H=a_1\sigma_1+a_3\sigma_3$ where $C_{2x}$ requires that  $a_{1,3}$ be even in $k_y$. Then, to lowest order, $a_{1,3}=M_{1,3x}k_x+M_{1,3y}k_y^2+N_{1,3}m$. Explicitly,
\begin{align}
\begin{pmatrix}
a_1\\ 
a_3
\end{pmatrix}
=&
\begin{pmatrix}
M_{1x}&M_{1y}\\ 
M_{3x}&M_{3y}
\end{pmatrix}
\begin{pmatrix}
k_x\\ 
k_y^2
\end{pmatrix}
+
m\begin{pmatrix}
N_{1}\\ 
N_{3}
\end{pmatrix} \nonumber \\
=&\textbf{M}\tilde{\textbf{k}}+m\textbf{N}.
\end{align}

Note that $\tilde{\textbf{k}}=(k_x,k_y^2)$. Using the QR decomposition, we can write $\textbf{M=\textbf{Q}\textbf{R}}$, where $\textbf{Q}$ is orthogonal and $\textbf{R}$ is upper triangular. Rewrite the Hamiltonian as 
\begin{align}
H=\sum_{i=1,2}{\sigma_i a_i}=& \sum_{i,j,k=1,2}{\sigma_j Q_{ji} (Q^T)_{ik}a_k} \nonumber \\ =& \sum_{i=1,2}{\sigma_i'a_i'},
\end{align}
where we have defined $\sum_{j=1,2}{\sigma_jQ_{ji}}=\sigma_i'$ and $\sum_{k=1,2}{(Q^T)_{ik}a_k=a_i'}$. Noting that $\textit{\textbf{a}}'=\textbf{Q}^T\textit{\textbf{a}}=\textbf{R}\tilde{\textbf{k}}+m\textbf{Q}^T\textbf{N}=\textbf{R}\tilde{\textbf{k}}+m \textbf{N}'$, the Hamiltonian takes the form
\begin{equation}
H=(R_{11}k_x+R_{12}k_y^2+mN_1')\sigma_1'+(R_{22}k_y^2+mN_2')\sigma_3'.
\end{equation}
If we make the correspondence $k_1\sim k_x$, $k_2\sim k_y$, this has the form of Eqn.~\eqref{sti1}. Setting $m=0$ takes us to \eqref{sti2}. Then, we see that this describes the evolution of a pair of Weyl points symmetrically with respect to the high symmetry lines $k_y=0,\pi$. 

\subsection{\thesubsection. \textbf{$\mathbf{C_3}$ and $\mathbf{C_{2x}}$ or $\mathbf{M_1}$}}
\addtocounter{subsection}{1}

In this subsection, we similarly discuss the case labelled \textbf{3l} in the main text. The groups \textbf{68}, \textbf{70}, \textbf{76}, and \textbf{78} contain 3-fold rotation about z axis and twofold rotation or mirror about in-plane axis as symmetries at K point. As shown in section SII S3, it is possible to create three pairs of Weyl points that evolves from K (KA) point. When this occurs, the representation for $C_3$ is $-\sigma_0$ and the representation for $C_{2x}$ ($M_1$) is $\pm i\sigma_3$. 

To describe this gap closing process, it is convenient to use polar coordinates $(r,\theta)$ with K point at $r=0$. We may also orient our axis so that $\theta=0$ corresponds to one of the high symmetry lines. As before, we demand that gap closes at $m=0$ while it stays open for $m<0$. The symmetries of the system implies that $H(r,\theta)=H(r,\theta+2\pi/3)$ and $H(r,-\theta)=-a_1(r,\theta)\sigma_1-a_2(r,\theta)\sigma_2+a_3(r,\theta)\sigma_3$. The former shows that we can Fourier expand in $\theta$ while the latter shows that $a_1,~a_2$ are odd and $a_3$ is even in $\theta$. Expanding the Hamiltonian to lowest order, $H=u_{1}'r^3\sin3\theta\sigma_1+u_{2}'r^3\sin3\theta\sigma_2+(u_{3m}m+u_{3r}r^2)\sigma_3$. Note that the analyticity of the Hamiltonian demands that $\sin3\theta$ should appear with $r^3$. After performing a rotation in the $\sigma_1$, $\sigma_2$ space, we may simplify the Hamiltonian as follows
\begin{equation}
H=u_{1}r^3\sin3\theta\sigma_1+(u_{3m}m+u_{3r}r^2).
\end{equation}
Finally, imposing the constraint that there is no gap closing for $m<0$, we get the constraint $u_{3m}u_{3r}<0$.  This describes closing of the gap and formation of 3 paris of Weyl points as shown in FIG.~\ref{SMgraphs} (a), (e), (j). When $m=0$, the Hamiltonian is 
\begin{equation}
H=u_{1}r^3\sin3\theta\sigma_1+u_{3r}r^2\sigma_3.
\end{equation}
The dispersion is quadratic in all directions as can be seen in FIG.~\ref{SMgraphs} (j).

\subsection{\thesubsection. $\mathbf{M_z}$}
\addtocounter{subsection}{1}

Finally, we discuss the case labelled by \textbf{loop} in the main text. This corresponds to the case when the eigenvalues of $M_z$ for the conduction and the valence bands are different, which restricts the Hamiltonian to $H=a_3 \sigma_3$. Expanding to first order, $a_3=b_1k_1+b_2k_2+a_mm$. The solution space of the gap closing condition is a plane in the parameter space, which is incompatible with the constraint that there is no solution for $m<0$. Thus, we include second order terms, $a_3=b_1k_1+b_2k_2+a_{11}k_1^2+2a_{12}k_1k_2+a_{22}k_2^2+a_mm$. The extremum for $a_3$ when $m=0$ must be $0$ at $k_x=k_y=0$. This condition for extremum gives $b_1=b_2=0$. If we now vary $m$, there should be a solution for $m>0$, and it must be a closed loop as we will be shown below. Assuming this for now, the solution must be an ellipse for small $m$. The condition for an ellipse is that $\det (\textit{\textbf{A}})>0$ where $\textit{\textbf{A}}$ is the matrix with components $a_{ij}$, $i,j=1,2$. Notice that we may diagonalize this matrix through an orthogonal matrix $P$. Defining $\textbf{k}'=P\textbf{k}$, $a_3=\lambda_1 k_1'^2+ \lambda_2 k_2'^2+a_mm$. The condition for ellipse now reads $\lambda_1 \lambda_2>0$ while the condition for solution coming into existence for $m\geq 0$ becomes $\lambda_1 a_m<0$ 

Now, we explain why the solution should be an ellipse. This is because a parabola requires $\lambda_1$ or $\lambda_2$ to be zero, which is not likely. On the other hand, a hyperbola would be in contradiction to our assumption because there would exist solution to the gap closing equation for arbitrary $m$. Thus, the Hamiltonian is
\begin{equation}
H=(\lambda_1 k_1'^2+ \lambda_2 k_2'^2+a_mm)\sigma_3.
\end{equation} 
This describes the gap closing and the formation of a line node as illustrated in FIG.\ref{SMgraphs} (a), (f), (k). When $m=0$, the Hamiltonian becomes
\begin{equation}
H=(\lambda_1 k_1'^2+ \lambda_2 k_2'^2)\sigma_3.
\end{equation}
Thus, the dispersion is quadratic in both directions.

\begin{figure*}[t]
\centering
\includegraphics[width=15cm]{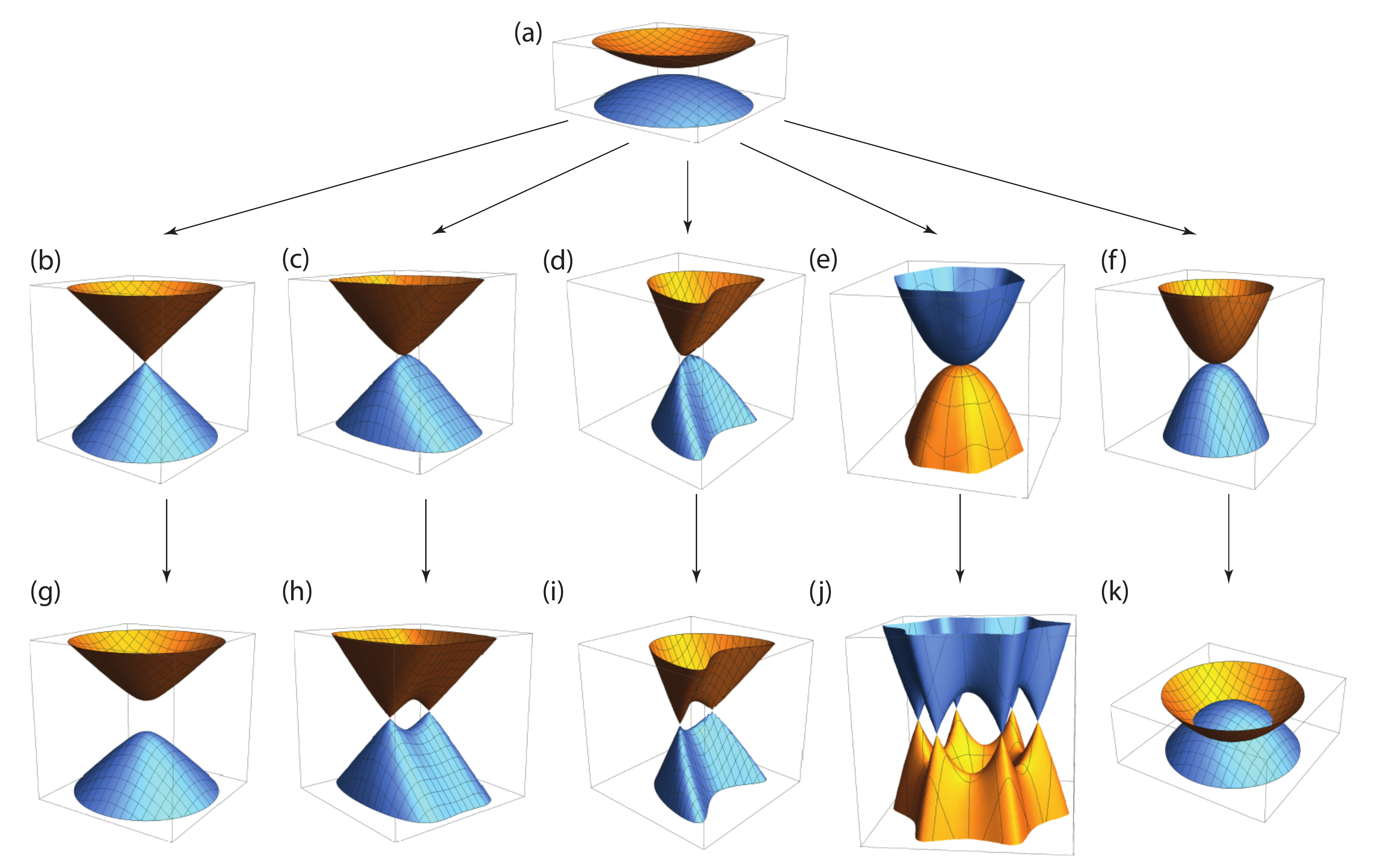}
\caption{Evolution of band structure near the critical point. (a) Generic dispersion before gap closes ($m<0$). (b) $\sim$ (f) Dispersion at the critical point ($m=0$) near the gap closing point. (g) $\sim$(k) Dispersion after gap closes ($m>0$).}
\label{SMgraphs}
\end{figure*}
\setcounter{subsection}{1}
\section{\thesection. Consideration of Time Reversal Symmetry}
\addtocounter{section}{1}

Although time reversal symmetry fixes only points in the Brillouin zone by itself, it may combine with other crystal symmetries to fix lines or planes. The former occurs when it combines with twofold rotation or mirror symmetry with in-plane axis while the latter occurs when it combines with twofold rotation with axis normal to the plane. We analyse the latter case first, then the former case, and finally, analyse high symmetry points that are not TRIM. Our goal will be to determine whether consideration of time reversal symmetry will induce extra double degeneracy, and if not, to determine whether there is any other possible emergent semimetallic phase which were not discussed in detail in the main text. In particular, we discuss the case labelled \textbf{3l} in the main text.

\subsection{\thesubsection. \label{plane} \textbf{High Symmetry Plane with Time Reversal Symmetry}} 
\addtocounter{subsection}{1}

In this subsection, we will prove that $C_{2z}\theta=I_{ST}$ can be represented by $K$ where $K$ is the complex conjugation operator. Because $C_{2z}$ is unitary and $\theta$ is antiunitary, $I_{ST}$ must be antiunitary. Thus, $I_{ST}=UK$ where $U$ is an N by N unitary matrix,
\begin{equation} 
UU^\dagger=1.
\label{cond1} \end{equation}
Also, the condition $I_{ST}^2=1$ implies that 
\begin{equation}
UU^*=1.
\label{cond2}\end{equation} 
These two conditions imply that $U^{-1}=U^\dagger=U^*$. Thus, $U$ is symmetric and unitary and we may write $U=e^{iM}$ where $M$ is symmetric and Hermitian. In other words, $M$ is real symmetric matrix, and such matrices can be diagonalized by a real orthogonal matrix. Since U transforms under real orthogonal change of basis by matrix $O$ as $U\rightarrow OUO^T$, we see that M can be diagonalized to a matrix $\phi$ with diagonal entries $\phi_{n}$, $n=1,...,N$. Then, $U$ is transformed to a diagonal matrix $e^{i\phi}$ with diagonal entries $e^{i\phi_n}$, $n=1,...,N$. Another transformation with matrix $D=diag(e^{-i\phi_1/2},...,e^{-i\phi_n/2})$ gets rid of the phase factors: $e^{i\phi}K\rightarrow De^{i\phi}KD^\dagger=K$. Thus, for any set of bands, $I_{ST}$ can be diagonalized, and we may discuss $I_{ST}$ acting on a single energy band (i.e. it does not introduce degeneracy). This means that we may talk about $I_{ST}$ acting on an arbitrary pair of bands as complex conjugation. 

\subsection{\thesubsection. \label{lines} \textbf{High Symmetry Line with Time Reversal Symmetry}} 
\addtocounter{subsection}{1}

The analysis for $I_{ST}$ can be applied whenever time reversal is combined with a unitary operator that commutes with it and squares to $-1$. It is then clear that $C_{2x}\theta$ and $M_y\theta$ also does not enforce double degeneracy. The same comment applies when $C_{2z}$ is replaced by $\{C_{2z}|ab\}$ because $\{C_{2z}|ab\}^2=-1$. The analysis becomes more complicated for $\{C_{2x}|ab\}\theta$ and $\{M_y|ab\}\theta$ where $a,~b=0$ or $1/2$. If we denote either of the operators by $R$, $R^2=-e^{i2ak_x}$ and  $(R\theta)^2=e^{i2ak_x}$. Writing $\{C_{2x}|ab\}\theta=UK$, we have in addition to \eqref{cond1}
\begin{equation}
UU^*=e^{i2ak_x}.
\label{cond3}\end{equation}
If $a=0$, the previous analysis applies and there is no degeneracy along the symmetry lines for $\{C_{2x}|ab\}\theta$, namely, the lines $k_x=0,~\pi$. In addition, because the basis can be chosen so that $R\theta=K$, gap closing event is not protected along the symmetry line ($H=a_1\sigma_1+a_3\sigma_3$ so two equations need to be satisfied for gap to close while there are two parameters, m and the momentum along the symmetry line). Since gap closing is not protected off the symmetry line, this does not lead to stable semimetallic phase. On the other hand, if $a=1/2$, $(R\theta)^2=-1$ so there is a double degeneracy along the line $k_x=\pi$, but not along the line $k_x=0$. \footnote{It can be shown~\cite{Weinberg} that if an antiunitary operator $A$ satisfies $A^2=e^{i\phi}$, it can be diagonalized if $e^{i\phi}=1$ while it can only be block diagonalized with 2 by 2 matrices of the form $\cos\frac{\phi}{2}\sigma_1-\sin\frac{\phi}{2}\sigma_2$ along the diagonal if $e^{i\phi}\neq 1$.}

Next, consider the possibility of multiple antiunitary symmetry along a line. This happens when there is a simultaneous presence of $C=\{C_{2x}|ab\}\theta$ and $M=\{M_y|a'b'\}\theta$ along the lines $k_x=0$ or $\pi$. If $a$ or $a'$ is $1/2$, there will be a double degeneracy along the lines as we have shown above. If we exclude these cases, they can be diagonalized individually but it is not clear if they can be simultaneously diagonalized. If we set $a=a'=0$, $CM:(x,y,z,t) \rightarrow(x,y+b-b',-z,t)\otimes(-i\sigma_3)$ and $MC: (x,y,z,t)\rightarrow (x,y-b+b',-z,t)\otimes(i\sigma_3)$, $CM=-e^{2ik_y(b-b')}MC$. Thus, along the symmetry lines, they either commute or anticommute. Writing $C=U_1 K$ and $M=U_2 K$, with symmetric and unitary $U_1$ and $U_2$, this condition becomes 
\begin{equation}
U_1U_2^*=\pm U_2U_1^*.
\label{cond4}
\end{equation}
Now, we showed above that $U_1=1$ with a suitable choise of basis so \eqref{cond4} implies that in this basis, $U_2$ is either real or purely imaginary. Because $U_2$ is symmetric and either $U_2$ or $iU_2$ is real, it can be diagonalized by real orthogonal transformation, under which $U_1$ will remain invariant. Thus, $C$ and $M$ can be simultaneously diagonalized. This analysis could have been carried out by considering the eigenvalues of $M_z$ since $MC\sim M_z$, but this clarifies how the two antiunitary operators can be simultaneously diagonalized. Note that stable semimetallic phase arise only when $R_c\neq R_v$ for the $M_z$ eigenvalues, which leads to a nodal line, as we already have seen.

We next consider the case when $I_{ST}=C_{2z}\theta$ is present with a nonsymmorphic rotation or mirror with in-plane axis where the translational part is non-zero for the direction normal to the line preserved by the rotation or mirror. In other words, the nonsymmorphic symmetries are of the form $\{C_{2x}|a b\}$ and $\{M_{y}|ab\}$ where $a=0$ or $1/2$ and $b=1/2$. We first note the action of $I_{ST}$ and $\{C_{2x}|a b\}$ on real space and spin space: $I_{ST}:(x,y,z,t)\otimes\sigma_0 \rightarrow (-x,-y,z,-t)\otimes i\sigma_1 K$ and $\{C_{2x}|a b\}:(x,y,z,t)\otimes\sigma_0\rightarrow(x+a,-y+b,-z,t)\otimes i\sigma_1$. Thus, $\{C_{2x}|a b\}I_{ST}:(x,y,z,t)\otimes\sigma_0\rightarrow(-x+a,y+b,-z,-t)\otimes(-K)$ and $I_{ST}\{C_{2x}|a b\}:(x,y,z,t)\otimes \sigma_0 \rightarrow(-x-a,y-b,-z,-t)\otimes K$. Thus, 
\begin{align}
\{C_{2x}|a b\}I_{ST}=& -T_{2a,2b}I_{ST}\{C_{2x}|a b\} \nonumber \\
=&-e^{i(2ak_x-2bk_y)}I_{ST}\{C_{2x}|a b\}.
\label{commutation} 
\end{align}
Here, $T_{2a,2b}$ is the translation operator with translation in x and y direction by $2a$ and $2b$ respectively. 

Now, we examine if $I_{ST}$ doubles the dimension of the representation by examining how the eigenvalue of the nonsymmorphic operator. Since $(\{C_{2x}|a b\})^2=-e^{2iak_x}$, the eigenvectors are $|\pm\rangle$ with eigenvalues $\pm i e^{iak_x}$. The question is whether $I_{ST}|\pm\rangle$ has the same $\{C_{2x}|a b\}$ eigenvalues. Using \eqref{commutation}, $\{C_{2x}|a b\}I_{ST}|\pm\rangle=\pm i e^{i(ak_x-2bk_y)}I_{ST}|\pm\rangle$. Now, it is easy to see that the eigenvalues switch iff $b=1/2$ and $k_y=\pi$ mod $2\pi$. The analysis for mirror symmetry is similar. See, for example, groups \textbf{20}, \textbf{21}, \textbf{24}, \textbf{25}. In hindsight, we see that this double degeneracy is actually due to $\{C_{2x}|ab\}\theta$ and $\{M_y|ab\}\theta$ where $a,~b=1/2$ along $k_x=\pi$ but the proof of the double degeneracy is simpler here due to the presence of unitary symmetry whose eigenvalues switch under the action of an antiunitary symmetry. Finally, note that when there is no double degeneracy, the stable semimetallic phase that may arise corresponds to the pattern ii:ij;\textbf{1s},\textbf{1l} discussed in the main text. This concludes the analysis of all subtleties that may arise along symmetry lines due to time reversal symmetry.

\subsection{\thesubsection. \label{points} \textbf{High Symmetry Points That Are Not TRIM}} 
\addtocounter{subsection}{1}

It is well known that time reversal forces double degeneracy at TRIM and we may exclude these points from our analysis. This leaves us with only K and KA points in the hexagonal Brillouin zone in Fig.~\ref{brillouin}. There are two questions that need to be addressed. Are there cases when there is no 1-D representation at K or KA? If not, can there be creation of stable band degeneracy starting from K or KA point by tuning an external parameter? The answer to the first question is no, as analysis of the inversion asymmetric groups show. The answer to the second question is yes. 

We tackle the second question first because this will answer much of the first question. To determine whether stable band degeneracy can evolve from K point, it helps to notice that protection of Weyl points is due to either $I_{ST}$ or $C_{2x}$ ($M_{y}$) type of symmetries when the Weyl points move off the symmetry point.

\subsubsection{K point in the presence of $I_{ST}$}
This requires the presence of 6-fold rotational symmetry in the crystal because there is both $C_{2z}$ and $C_3$ symmetry. We present the analysis for group \textbf{73} which contains only 6-fold rotation in addition to translations. The expectation that gap closes at K and that $I_{ST}$ will protect the subsequent creation of 3 pairs of Weyl points is not met.

We showed previously that $I_{ST}$ may be represented by the complex conjugation operator $K$. On the other hand, presence of additional symmetry such as $C_3$ can complicate matters because in the representation where $I_{ST}=K$, $C_3$ is not in general a diagonal matrix despite the fact that $C_3$ and $C_{2z}\theta$ commute (because $I_{ST}$ is antiunitary). In fact, operation of $C_3$ may mix states between different bands so it may not even be possible to talk about $C_3$ with arbitrary pair of bands (because the action of $C_3$ will take states in one of these two bands into a state from a different band). 

To make this clear, begin by finding the eigenvalues of the operator $C_3$. Since $(C_3)^3=-1$, the eigenvalues are $e^{i(\pi/3+2\pi n/3)}$ where n is an integer. If it were to be possible to talk about an arbitrary pair of bands so that the pair of bands have $C_3$ and $I_{ST}$ symmetries, it must be possible to choose representations for these operators so that $I_{ST}=K$ and $C_3$ has two arbitrary eigenvalues that cubes to $-1$. We will show below that this is impossible. This implies that an arbitrary pair of energy bands will not simultaneously host $I_{ST}$ and $C_3$ because both of these are symmetries at K point in the Brillouin zone. 

As shown before, we may take $I_{ST}=\sigma_0K$ for any pair of bands. We find the possible representation for $C_3$ for arbitrary pair of bands under the constraint that the action of $C_3$ does not take us to states outside those in the two bands. The most general form of $C_3$ is
\begin{equation}
C_3=\sum_{i=0}^3{c_i\sigma_i}.
\end{equation}
Next, impose the following constraints
\begin{equation}
[C_3,I_{ST}]=0, ~(C_3)^3=-1,~C_3C_3^\dagger=1.
\end{equation}
We tackle one constraint at a time

(i) $[C_3,I_{ST}]=0$: It is easy to see that this condition implies that $c_0,~c_1,~c_3$ are real while $c_2$ is purely imaginary.

(ii) $(C_3)^3=-1$: Denoting by $\vec{c}=(c_1,c_2,c_3)$, short calculation shows that this gives $c_0^3+3c_0\vec{c}^{~2}=-1$ and $3a_0^2+\vec{c}^{~2}=0$

(iii) $C_3C_3^\dagger$: This gives three constraints, $c_0^2+c_1^2+c_3^2-c_2^2=1$, $c_0c_1+ic_2c_3=0$, and $c_0c_3-ic_1c_2=0$.

It follows from $c_0c_1+ic_2c_3=0$ and $c_0c_3-ic_1c_2=0$ that $c_1=0$ or $c_0^2=c_2^2$. If $c_1=0$, the same two conditions show that either $c_3=0$ or $c_0=c_2=0$. The latter is impossible because $c_0^2+c_1^2+c_3^2-c_2^2=1$ shows that $c_3^2=1$ while $3a_0^2+\vec{c}^{~2}=0$ shows that $c_3^2=0$. On the other hand, if $c_0^2=c_2^2$ shows that $c_0=c_2=0$ because $c_0$ is real while $c_2$ is purely imaginary. The remaining conditions $3a_0^2+\vec{c}^{~2}=0$ and $c_0^2+c_1^2+c_3^2-c_2^2=1$ cannot be simultaneously true. Thus, the only possibility is that $c_1=c_3=0$. 

If $c_1=c_3=0$, the remaining two conditions are  $c_0^3+3c_0c_2^2=-1$ and $3c_0^2+c_2^2=0$. If $c_2=0$, $c_0=-1$ while if $c_2\neq 0$, $c_0=\frac{1}{2}$ and $c_2=\pm i \frac{\sqrt{3}}{2}$.

In conclusion, if $I_{ST}=K$, there are only three possibilities: $C_3=-\sigma_0,~\frac{1}{2}\sigma_0 \pm i \frac{\sqrt{3}}{2}\sigma_2$. Thus, the only allowed pairing of $C_3$ eigenvalues are $\{-1,-1\}$ and $\{e^{i\pi/3}, e^{-i\pi/3}\}$. In the former case, $C_3$ does not constrain the form of the Hamiltonian at K point while in the latter case, the two bands are doubly degenerate. 

To summarize, suppose that we choose two arbitrary bands. We have shown that it is possible to choose $I_{ST}=K$. However, whether we can speak of $C_3$ symmetry acting on these two bands depend on the  eigenvalues of $C_3$ at K point. If it is possible to speak of $C_3$, the eigenvalues of the two bands must be paired as $\{-1,-1\}$ or $\{e^{i\pi/3}, e^{-i\pi/3}\}$. Otherwise, we must add two additional energy bands to get a 4-band model to speak of $C_3$ operator.

We note that this can be seen in a different way by examining how the $C_3$ eigenvalue of a state change under the operation of $I_{ST}$.   Denote a state having $C_3 $ eigenvalue $e^{i(\pi/3+2\pi n/3)}$ by $|n\rangle$. Then $I_{ST}|n\rangle$ has eigenvalue $e^{-i(\pi/3+2\pi n/3)}$. This means that unless the eigenvalue is $-1$, $I_{ST}$ imposes double degeneracy. Also, if we want to talk about $C_3$ and $I_{ST}$ simultaneously on a two-band model, the eigenvalues must be paired as $\{-1,-1\}$ or $\{e^{i\pi/3}, e^{-i\pi/3}\}$, in agreement with the previous analysis.

\subsubsection{K point in the presence of $C_{2x}$ or $M_{1}$ type of symmetry}
The simplest case is when there is only the threefold rotation and $C_{2x}$ or $M_1$ (twofold rotation or mirror whose symmetry axis passes through K point) as in groups \textbf{68} and \textbf{70} respectively. The 1-D representation for $C_3$ is $-1$ while those for twofold rotation or mirror is $\pm i$. This is due to the relation $C_3P=PC_3^{-1}$ where $P$ is either $C_{2x}$ or $M_{y}$ which implies that unless a state has eigenvalue $-1$ for $C_3$, the representation cannot be one dimensional. 

For two pairs of energy bands whose $C_3$ eigenvalue is $-1$ at K point, it is the eigenvalues of $P$ that determine whether bands may close at the high symmetry point. If $R_c=R_v$ for $P$, gap does not close at K point in general because $P\propto \sigma_0$ but it may close if $R_c\neq R_v$ because $P\propto \sigma_3$. After gap closes, there will be evolution of three pairs of Weyl points along the three high symmetry lines that cross at K point because $R_c\neq R_v$ along these lines and the problem reduces to \textbf{1l} discussed in the main text. This pattern of gap closing is labelled \textbf{3l}. Note that the mechanism for protection of Weyl points in this case is the same as that for $C_{2x}$ or $M_y$. 

Next, we discuss the case with $C_3$ replaced by $C_6$ symmetry, which is equivalent to considering an additional $I_{ST}$ symmetry at K point. This occurs for \textbf{76} and \textbf{77}, which contain $C_{2x}$ or $M_{1}$ respectively in addition to $C_3$ and $I_{ST}$ at K point. From the above analysis, the only 1D representation possible is $C_3=-1$ and $P=\pm i$ where $P=C_{2x}$ or $M_1$. The claim is that $I_{ST}$ does not force degeneracy. This is easy to see because we have already shown in section SII S2 that the group relation between $I_{ST}$ and $P$ is consistent with the representation, and we have shown in the previous subsection that the group relation between $I_{ST}$ and $C_3$ is consistent with the representation, and finally, we have shown in this subsection that the group relation between $C_3$ and $P$ is consistent with the representation. The conclusion follows by observing that $C_3$, $P$, and $I_{ST}$ generates the group. 

Now, \textbf{76} contains \textbf{68} as subgroup and \textbf{77} contains \textbf{70} as subgroup. Restricting the representation for \textbf{76} and \textbf{77} to these subgroups, we obtain the representation for \textbf{68} and \textbf{70} that was found previously. Thus, we see then that if $R_c=R_v$ for $P$, gap does not close at K while if $R_c\neq R_v$, it is possible to obtain \textbf{3l}.

\subsubsection{Possibility of Additional Double Degeneracy at K point}

Now, we come back to the question of whether consideration of time reversal symmetry can forbid one dimensional representation at K point. We begin by listing all of the possible symmetries
\begin{equation}
C_3, M_2, C_{2x},M_z, M_x\theta, C_{22}\theta, I_{ST}.
\end{equation}
Here, $M_{2}$ and $C_{2x}$ is a mirror symmetry or twofold rotation symmetry that leaves invariant one of the high symmetry line passing through the K point (They fix the line LE passing through the KA point in Fig.~\ref{brillouin} (e)). Note that we have not listed symmetries that can be formed by combining one of the symmetries we have listed with $C_3$, which will always be present for hexagonal Brillouin zone. For example, $M_1$ and $M_3$, which are also mirror symmetries that leaves invariant one of the high symmetry lines passing through the K point, are not listed because they can be obtained by a suitable combination of $M_2$ and $C_3$. Note also that $C_{22}$ is a twofold rotation symmetry and $M_x$ is a mirror symmetry fixing the line $SN$ in Fig.~\ref{brillouin} (e). It can be shown by going through all of the combinations that it suffices to consider only the following symmetries in addition to $C_3$ at K point.
\begin{equation}
(M_x\theta), (C_{22}\theta), (I_{ST}), (M_1,I_{ST}), (C_{2x},I_{ST}), (C_{22}\theta,M_x\theta).
\end{equation}
Note that the combination is such that there is no inversion symmetry and time reversal symmetry appears in combination with some spatial symmetry. Before moving on, we note that there does exist one case where there is no 1D irrep because of the simultaneous presence of $M_2$ and $C_{2x}$, which has been discussed in the main text (see group \textbf{79}). 

We have actually carried out most of the calculations needed to determine that addition of time reversal symmetry to the system does not prohibit 1D representation at K point. The presence of  $C_{22}\theta$ or $M_x\theta$ in addition to  $C_3$ appears as subgroup of \textbf{76} and \textbf{77} respectively. This leaves us with the case with $C_{22}\theta,~M_x\theta$. However, we have already shown that it is possible to simultaneously diagonalize these symmetries, which means it is possible to talk about these symmetries acting on one band. In general, we may take $C_{22}\theta=K$, $M_x\theta=\pm i K$ to satisfy \eqref{cond4}. A candidate representation for $C_3$ is $-1$. We have shown that this representation is consistent with the group relation between $C_{22}\theta $ and $C_3$. If $M_x\theta=K$, our previous calculation would show that this is also be consistent with $C_3=-1$. However, phase factor in front of $K$ is irrelevant for the group relation between $M_x\theta$ and $C_3$. This concludes the proof.

\setcounter{subsection}{1}
\section{\thesection. Topological Charge}
\addtocounter{section}{1}

In this section, we show that the emergence of stable band degeneracy is always accompanied by a (quantized) topological charge. We define topological charge for each of the three classes of symmetry.

(i) $C_{2z} \theta$: Under time reversal symmetry, the Berry curvature satisfies $\Omega (-\textbf{k})=\Omega(\textbf{k})$ while under the rotation symmetry, $\Omega(-\textbf{k})=\Omega(\textbf{k})$. Thus, $\Omega (\textbf{k})=-\Omega (\textbf{k})$ under $C_{2z} \theta$ and the Berry curvature vanishes everywhere except for singularities realized by Weyl points \cite{SFang-Fu}. This quantizes the Berry phase in units of $\pi$. 

(ii) $\{M_{z}|\textbf{t}\}$: First, note that eigenvalues are $\pm c$, $|c|=1$. Following \cite{topological}, pick a point $\textbf{k}_1$ `inside' the loop and another point $\textbf{k}_2$ `outside' the loop. Define $N_{\pm}(\textbf{k})=N_{\pm}^c(\textbf{k})-N_{\pm}^v(\textbf{k})$. Here, $N^{c(v)}_{\pm}(\textbf{k})$ is the number of conduction (valence) bands with eigenvalues $\pm c$ at $\textbf{k}$. The charge is defined to be (see Fig.~\ref{charge})
\begin{equation}
Q=\frac{\pi}{4} [N_{+}(\textbf{k}_1)-N_{-}(\textbf{k}_1)-N_{+}(\textbf{k}_2)+N_{-}(\textbf{k}_2)].
\label{topcharge}
\end{equation}

(iii) $\{C_{2x}|\textbf{t}\}$ (or $\{M_{y}|\textbf{t}\}$): The charge is defined exactly as in \eqref{topcharge} but with $\textbf{k}_1$ and $\textbf{k}_2$ along the symmetry axis with $\textbf{k}_1$ to the left and $\textbf{k}_2$ to the right of the gap-closing point. We note that the topological charge can also be defined by integrating along a curve symmetric with respect to the symmetry line. In this case, $C_{2x}$ (or $M_y$) implies that $\Omega(k_x,k_y)=-\Omega(k_x,-k_y)$ so the integral vanishes unless there is a singularity.
 
\begin{figure}[ht]
\centering
\includegraphics[width=7cm]{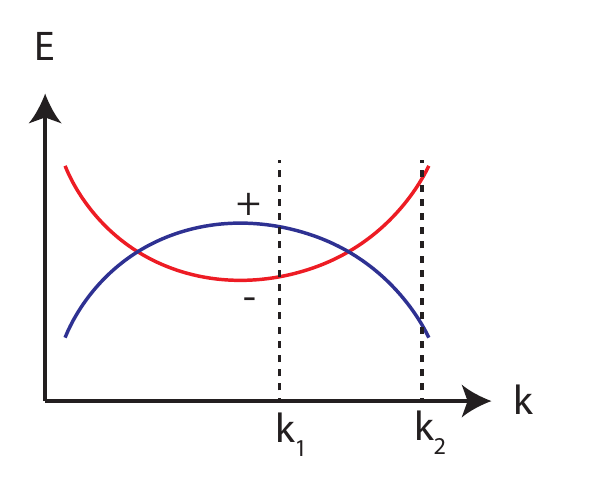}
\caption{Schematic diagram of band crossing for case (ii) and (iii). In case (ii), k axis is along the line connecting $k_1$ and $k_2$, while in case (iii) it is along the symmetry line. The blue (red) band has eigenvalue $+c$ ($-c$). For the setup in the figure, the charge is $\pi$ }
\label{charge}
\end{figure}
\setcounter{subsection}{1}
\section{\thesection. Black Phosphorous}
\addtocounter{section}{1}

In this section, we present a simple application to the kp model of black phosphorous. As shown in \cite{Ahn-Yang}, the kp Hamiltonian near the $\Gamma$ point takes the form
\begin{widetext}
\begin{equation}
H(k_x,k_y)=Ak_x\sigma_y+(M-B_1k_x^2-B_2k_y^2)\sigma_z+\lambda_1s_y\sigma_y+\lambda_2k_ys_z\sigma_x.
\end{equation}
\end{widetext}
Here, $s_i$ and $\sigma_i$ are the Pauli matrices for spin and orbital degrees of freedom, respectively, M is a tunable parameter, and $A,~ B_1,~ B_2,~ \lambda_1,~ \lambda_2$ are constants. Black phosphorous has puckered structure and when the symmetry is lowered by breaking the inversion symmetry, it belongs to layer group \textbf{24}, which contains $M_x$, and $M_y$ symmetries (note that $M_x M_y \sim C_{2z}$). Although the mirror symmetries are nonsymmorphic, this is irrelevant for kp theory near the $\Gamma$ point. Taking this into account, the symmetries take the following representations
\begin{equation}
M_x=is_x\sigma_z,~M_y=is_y,~\theta=is_yK.
\end{equation}
Here, $\theta$ is the usual time reversal symmetry. As we tune the parameter $M$ , the gap may close or open along $k_x=0$ or $k_y=0$. Our claim is that this gap closing follows the pattern \textbf{1s}. As an example, we verify this along $k_y=0$ along which $M_y$ is a symmetry. In particular, we show that the $M_y$ eigenvalues for the gap closing bands are equal. To do this, set $k_y=0$ in the Hamiltonian to get
\begin{equation}
H(k_x,k_y)=Ak_x\sigma_y+(M-B_1k_x^2)\sigma_z+\lambda_1s_z\sigma_y.
\end{equation}
Notice that we have changed the basis in the spin sector so that $s_y\rightarrow s_z$. In this basis, $M_y=is_z$. Now, it is easy to see that the gap closes between bands in the sector with the same $s_z$ eigenvalues, which means that the $M_y$ eigenvalues are equal for the bands that cross.
\setcounter{subsection}{1}
\section{\thesection. Brillouin Zone}
\addtocounter{section}{1}

For the reader's convenience, we have organize the layer groups according to their Brillouin zone in Table~\ref{LG} and illustrated the Brillouin zone in Fig.~\ref{brillouin} with the convention used by Litvin and Wike \cite{SLitvin}. 

\begin{table*}[ht]
\centering
\caption{LG}
\label{LG}
\begin{tabular}{|l|l|}
\hline
Brillouin zone & Layer group\\
Oblique p & \textbf{1}, \textbf{3}, \textbf{4}, \textbf{5}\\
Rectangular p & \textbf{8}, \textbf{9}, \textbf{11}, \textbf{12}, \textbf{19}, \textbf{20}, \textbf{21}, \textbf{23}, \textbf{24}, \textbf{25}, \textbf{27}, \textbf{28}, \textbf{29}, \textbf{30}, \textbf{31}, \textbf{32}, \textbf{33}, \textbf{34}  \\
Rectangular c & \textbf{10}, \textbf{13}, \textbf{22}, \textbf{26}, \textbf{35}, \textbf{36}, \\
Square p & \textbf{49}, \textbf{50}, \textbf{53}, \textbf{54}, \textbf{55}, \textbf{56}, \textbf{57}, \textbf{58}, \textbf{59}, \textbf{60}\\
Hexagonal p & \textbf{65}, \textbf{67}, \textbf{68}, \textbf{69}, \textbf{70}, \textbf{73}, \textbf{74}, \textbf{76}, \textbf{77}, \textbf{78}, \textbf{79}\\
\hline

\hline
\end{tabular}
\end{table*}

\begin{figure*}[ht]
\centering
\includegraphics[width=17.5cm]{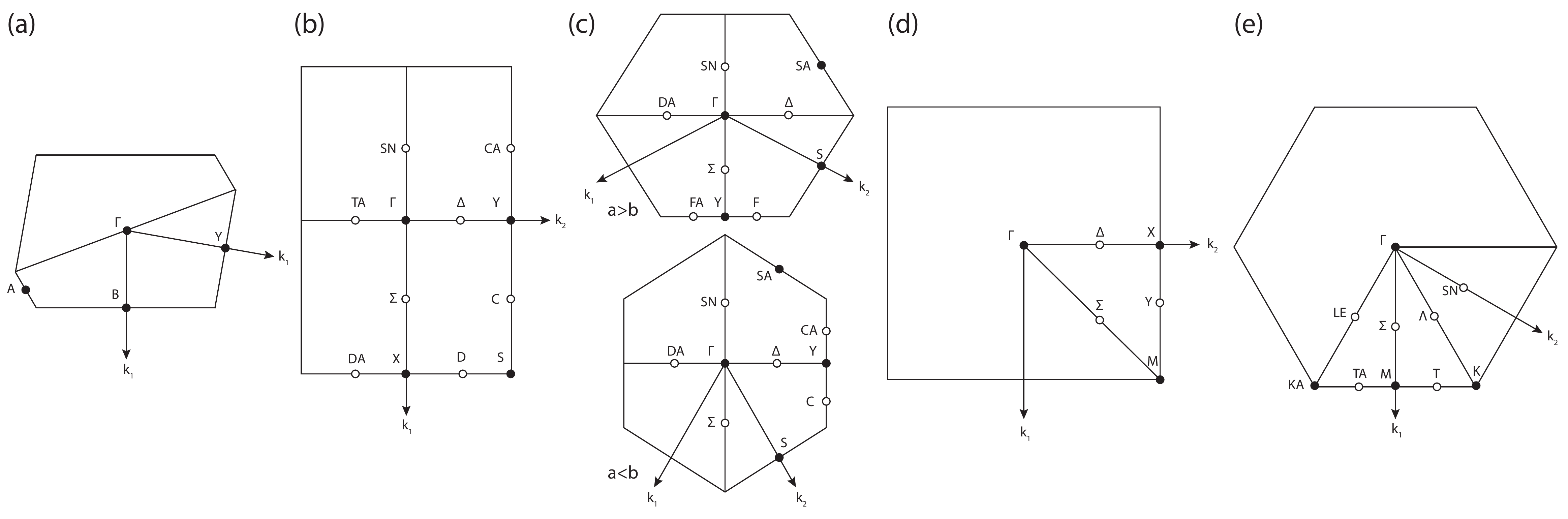}
\caption{(a) Primitive oblique (b) Primitive rectangular (c) Centered rectangular (d) Primitive square (e) Primitive hexagonal}
\label{brillouin}
\end{figure*}


\end{document}